\documentstyle[dina4,12pt]{article}

\begin{document}
\renewcommand{\theequation}{\arabic{section}.\arabic{equation}}
\parskip=4pt plus 1pt
\textheight=8.7in
\textwidth=6.0in
\newcommand{\be}{\begin{equation}}
\newcommand{\ee}{\end{equation}}
\newcommand{\bea}{\begin{eqnarray}}
\newcommand{\eea}{\end{eqnarray}}
\newcommand{\co}{\; \; ,}
\newcommand{\nn}{\nonumber \\}
\newcommand{\hlogm}{\ln \frac{M^2}{\mu^2}}
\newcommand{\apbc}{(\alpha + \beta)^C}
\newcommand{\ambc}{(\alpha - \beta)^C}
\newcommand{\apbn}{(\alpha + \beta)^N}
\newcommand{\ambn}{(\alpha - \beta)^N}
\newcommand{\ambcn}{(\alpha - \beta)^{C,N}}
\newcommand{\apbcn}{(\alpha + \beta)^{C,N}}
\newcommand{\hmpp}{M_{\pi}}
\newcommand{\mppq}{M_{\pi}^2}
\newcommand{\p}[1]{(\ref{#1})}
\newcommand{\D}[1]{{\cal D}^{#1}}
\newcommand{\gag}{$\gamma \gamma \rightarrow \pi^0 \pi^0$}
\newcommand{\gpgpz}{$\gamma\pi^0 \rightarrow \gamma \pi^0$}
\newcommand{\gpgp}{$\gamma\pi \rightarrow \gamma \pi$}
\newcommand{\ggppz}{$\gamma\gamma\rightarrow\pi^0\pi^0$}
\newcommand{\ggpp}{$\gamma\gamma\rightarrow\pi\pi$}
\newcommand{\mscript}[1]{{\mbox{\scriptsize #1}}}
\newcommand{\mtiny}[1]{{\mbox{\tiny #1}}}
\newcommand{\MS}{\mtiny{MS}}
\newcommand{\GeV}{\mbox{GeV}}
\newcommand{\MeV}{\mbox{MeV}}
\newcommand{\keV}{\mbox{keV}}
\newcommand{\ren}{\mtiny{ren}}
\newcommand{\kin}{\mtiny{kin}}
\newcommand{\hint}{\mtiny{int}}
\newcommand{\tot}{\mtiny{tot}}
\newcommand{\CHPT}{\mtiny{CHPT}}
\newcommand{\DISP}{\mtiny{DISP}}
\newcommand{\CA}{\mtiny{CA}}
\newcommand{\scs}{\co \;}
\newcommand{\sem}{ \; \; ; \;}
\newcommand{\per}{ \; .}
\newcommand{\la}{\langle}
\newcommand{\ra}{\rangle}
\newcommand{\unith}{{\bf{\mbox{1}}}}
\begin{titlepage}
\noindent
\vspace*{0.5cm}
\begin{center}
{{\Large{\bf{{Pion (Kaon) and Sigma polarizabilities}}
}}}$^\sharp$

\vspace{1cm}

{\large S. Bellucci}
\vspace{0.5cm}

INFN-Laboratori Nazionali di Frascati,
\\P.O.Box 13, 00044 Frascati, Italy\\

\vspace{1.2cm}

\end{center}
\begin{abstract}
\noindent
We report the results of the working group on ``Pion (Kaon) and Sigma
Polarizabilities''. Interesting possibilities
to measure these polarizabilities include the radiative pion photoproduction
in the MAMI experiment at Mainz, as well as at the GRAAL facility
(actually the latter is being considered for an experimental
determination of the pion polarizabilities here for the first time),
the experimental plans on Primakoff effect at FNAL, and the measurements
at the Frascati $\Phi$-factory DA$\Phi$NE.

\noindent
{\underline{\hspace{5cm}}}\\

\noindent
$\sharp$ Work supported in part by the EEC Human Capital and Mobility Program.
To appear in the proceedings of the Chiral Dynamics Workshop, 25-29 July 1994,
MIT.

\vspace{.15cm}

\noindent
e-mail: Bellucci@lnf.infn.it
\end{abstract}
\end{titlepage}
\newpage
\setcounter{page}{2}
\setcounter{section}{0}
\setcounter{equation}{0}
\setcounter{subsection}{0}

         \section{Introduction\label{in}}

	We report here about the activity of the working group
on ``Pion (Kaon) and Sigma Polarizabilities'' we conducted, in collaboration
with R. Baldini, at the Workshop on Chiral Dynamics:
Theory and Experiment, held July 25-29,1994 at MIT. The goal of
this working group was to identify the processes that are more suitable
to measure the electric and magnetic polarizabilities of the abovementioned
hadrons and probe chiral dynamics in the photon-pion (-kaon) and photon-sigma
physics.

	The agenda of the group was a mixture of theory and
experiment that allowed us to summarize the current status in this
field, determine what is to be done in order to improve it, from both
the theoretical and the experimental side, and quantify the level of
accuracy needed to make the improvement significant.

	We considered the following general areas:

1. Theoretical predictions and models for Compton scattering
   $\gamma\pi(K)\rightarrow\gamma\pi (K)$, as well as for
$\gamma\gamma\rightarrow\pi\pi (K K)$,
   and the relation to the pion (kaon) polarizabilities.
2. Experiments to measure the pion (kaon) and sigma polarizabilities.
3. Phenomenology required in order to extract the polarizabilities
   from the experimental data.

	We have considered the following methods of measurement:

i) Radiative photoproduction of the pion and extrapolating to the pion pole,
in order to extract the polarizabilities form the data.
ii) Experiments to measure the pseudoscalar meson polarizabilities using the
Primakoff effect.
iii) The measurements at the Frascati DA$\Phi$NE with the KLOE detector.

	Murray Moinester illustrated the plans at FNAL and the reaction
$\pi\rho\rightarrow\pi\gamma$ in connection with the pion and sigma
polarizabilities.
Thomas Walcher went over the MAMI project at Mainz. Annalisa D'Angelo
discussed the potential of the Graal synchrotron light facility in
Grenoble to carry out polarizability measurements.
S. Kananov discussed the need for a careful estimate of the
radiative corrections for the FNAL experiments where the pion
polarizabilities can be measured.
A presentation of the DA$\Phi$NE capabilities has been done
in the workshop.

	The related talks had the following titles:
A. D'Angelo: "The experimental plans in Grenoble",
T. Walcher: "The MAMI experiment at Mainz",
M. Moinester: "Pion polarizabilities and quark-gluon plasma signatures",
S. Kananov: "Radiative corrections for pion polarizability experiments",

	In the second part of the working group, devoted to the theoretical
contributions, we had the following talks:
J. Kambor: "Determination of a $O(p^6)$ counterterm from sum rules",
M. Knecht: "$\gamma\gamma\rightarrow\pi^0\pi^0$ and $\pi^0$ polarizabilities
in generalized chiral perturbation theory",
M. Pennington: "Dispersion relations and pion polarizabilities",
S. Bellucci: "Difficulties in extrapolating to the pion pole the data on
radiative pion photoproduction".

\setcounter{equation}{0}
\setcounter{subsection}{0}
\section{Experimental plans for polarizability measurements\label{exp}}

\subsection{Radiative pion photoproduction}

Let us consider first of all the unexpected and very innovative
contribution by Annalisa D'Angelo. She has reported about the
possibility to measure the pion polarizability by Graal, the new
facility at the electron storage ring ESRF in Grenoble.

The Graal facility consists of a tagged and highly polarized
$\gamma$-ray beam, produced by the backscattering of Laser light against the
high energy electrons circulating in the storage ring ESRF at
Grenoble \cite{z1}-\cite{z3}.

If commercial Ion-Argon and Nd-Yag Lasers are used, either linearly or
circularly polarized, a $\gamma$-ray beam of energy ranging from
about 300 MeV to 1.5 GeV with a degree of polarization higher then 70$\%$ over
almost the entire spectrum may be obtained, by appropriate choice of the
Laser line.

A large solid angle multi purpose detector
\cite{z4} is part of the Graal facility
and it will be used to perform experiments on photo-nuclear reactions
\cite{z5}.
It consists of a crystal ball made of 480 BGO crystals (24 cm long)
covering all
azimuthal angles for polar angles in the interval between 25$^{\circ}$ and
155$^{\circ}$. It may be used as electromagnetic calorimeter, with measured
energy resolution of 2$\%$ FWHM for 1 GeV photons, or to detect protons of
energy
up to 300 MeV \cite{z6}-\cite{z8}.

The central hole of the BGO ball ($\phi$ = 20 cm) will contain a barrel of 32
plastic scintillators; it will be used to discriminate between charged and
neutral particles and to identify the charged particles with the
$\frac{\Delta E}{E}$
technique. Inside the barrel two
cylindrical wire chambers will provide
improved angular resolution for the reconstruction of the trajectories of the
charged particles.

In the forward direction two detectors, each consisting of three plane wire
chambers rotated of 45$^{\circ}$, and a scintillating wall will provide charged
particle tracking information and TOF measurements; 10$\%$ efficiency is
expected for neutron detection in the scintillating wall.

We have started to investigate the possibility of using the Graal facility to
study the radiative pion photoproduction from the proton (namely the reaction
$\vec{\gamma} + p \rightarrow n + \pi^{+} + \gamma$), in order to extrapolate
the experimental data to the pion pole and determine the cross-section of the
Compton scattering on the pion \cite{z9}.

The interest of this measure, in order to get information on the
pion polarizability, has been pointed out, among other authors, by D.
Drechsel and L.V. Fil'kov \cite{z9};
they have stressed that in order to obtain a
reliable extrapolation it is necessary to have experimental data in kinematical
condition as close as possible to the point $t=0$, being $t$ the momentum
transferred between the final neutron and the initial proton.

A measurement performed at 1.5 GeV photon
energy, using polarized photons, would
fulfill the experimental requirements of Ref. \cite{z9}, also providing higher
sensitivity to the pion polarizability contributions through the polarization
structure functions.

A fundamental issue of the experimental set-up is the capability
of discriminating the reaction of interest from the background events, like
those coming from asymmetric decay of $\pi ^{0}$ in the $\pi^+ \pi^0 n$
reaction channel.

In principle all these requirements are fulfilled by the Graal facility:
the scattered photons may be detected in the BGO ball for laboratory angles
between $25^{\circ}$ and $155^{\circ}$; low energy  neutrons my be detected in
the forward direction using the plastic wall and they may be identified using
the TOF information; finally the pions may be detected at all
angles by the wire
chambers with good angular resolution. All these experimental information
should allow a complete reconstruction of the interesting events in selected
kinematical conditions, with expected good background rejection.

The Graal facility set-up is therefore a promising tool to perform the
first experiment with polarized photons on radiative pion
photoproduction in order to extract information on the pion
polarizability.
\\

Thomas Walcher has reviewed, from his general talk on
the experimental activity at MAMI in Mainz,
the measurement of the pion polarizability still by means of radiative pion
photoproduction.

 Walcher has shown some kinematical conditions suitable
for the measurement of the charged pion polarizability.
The sensitivity in the Chew-Low extrapolation at the pion pole has been
stressed \cite{z9}. For instance a variation of ${\alpha}_{\pi}$
from 0 up to 7 is equivalent to a 20 $\%$ variation in the extrapolated
amplitude. Hence the extrapolation has to be done at the
2 $\%$ level , if $\alpha_{\pi}$ has to be measured at a 10 $\%$
level. This sensitivity depends, of course,
on the minimum momentum transfer
$t_{min}$ achieved.
 For instance for a 700 MeV incident photon,
a 152 MeV final photon, a neutron in the angular range
$12^{0}$-$32^{0}$ and in the energy range 400 - 350 MeV, it is
$t$= 0.31, in pion mass squared units. Nucleon, pion and $\Delta(1236)$
pole diagrams have been evaluated and the extrapolation seems
feasible.

Conversely the measurement of the neutral pion polarizability is not realistic
at the MAMI energies. The situation may be improved at higher
 incident photon energies and a polarized beam would be very welcome,
just like the Graal facility!
By the way M. Moinester has stressed that the maximum energy
meaningful for extracting the pion polarizability is about 2 GeV,
 corresponding
to the $\rho$ mass in the photon-pion c.m. system.

The MAMI detector consists of a MWPC system to detect the charged
pion,  close to the incident
photon beam, a segmented $BaF_{2}$ to detect the scattered photon
and a system of scintillators for detecting the neutron  and
providing the neutron time of flight.The expected yield
of radiative pion photoproduction is
$\simeq$ 2000 events/day. The background due to double
photoproduction, simulating radiative pion photoproduction,
 is $\simeq$ 160 events/day.

\subsection{Primakoff effect}

The contribution of Murray Moinester has concerned the measurement of
 $\pi$ ($K$) and $\Sigma$ polarizabilities via $\pi$ ($K$) and $\Sigma$
high energy beams at Fermilab by E781 \cite{z10}.

 This experimental activity has been reported already
in detail by M. Moinester in his contribution to the workshop.
Therefore only the main topics are emphasized here, first of all
the relationship between polarizability and radiative transitions.

The $A_{1}$ radiative width is a good illustration of this
statement. It has been demostrated \cite{z11}
according to the current algebra the main
contribution to the pion electric polarizability comes from the exchange
of the  $A_{1}$. Xiong, Shuryak and Brown
\cite{z12} have
shown that a radiative width $\Gamma$($A_{1}\rightarrow\pi\gamma$)=
1.4 MeV is needed to get the current algebra value, on the
contrary the experimental value is $\Gamma$($A_{1}\rightarrow\pi\gamma$)=
0.64 $\pm$ 0.25 MeV \cite{z13}. More data from E781 are welcome to
settle this relevant problem.

By the way the unexpected correlation between the reaction
$\pi \rho \rightarrow \pi \gamma$ and the photon flux in a quark-gluon
plasma has been pointed out.

Experimental results from the previous FNAL experiment E272 have
been shown to demonstrate the experimental feasibility of the
Primakoff effect, even if E272 did not get enough statistics
to measure $\alpha_{\pi}$. The main experimental problem in
measuring the Primakoff effect by the new FNAL experiment, E781,
concerns a suitable fast trigger. It is not implemented at the
moment, taking into account the high rate and the difficulties
for using any signal from the scattered photon detector, which
is 50 meters downstream the target.

{}From a theoretical point of view an open question remains the
fair disagreement between the charged $\rho$ radiative width,
as obtained via the Primakoff effect, and the neutral $\rho$
radiative width.
\\

Finally M. Moinester has stressed the role of radiative transitions
in the case of a $\Sigma$ beam, which should also be available in E781.
The radiative transitions to the Sigma*(1385) provide a measure of the s-quark
magnetic moment of the Sigma\cite{lipkin}.
Positive and negative $\Sigma$ are expected to have very
different polarizabilities. In particular  the  $\Sigma^{-}$ magnetic
moment should
be negligible, both taking into account the 3 quarks have the
same charge and according to the U-spin symmetry. Furthermore
it has been
shown that the event rate for measuring $\Sigma$ radiative
transitions by a 600 GeV $\Sigma$ beam is higher than the
expected rate for measuring the pion Primakoff effect by
E781.
\\

S. Kananov has reported about radiative corrections in the
scattering of pions by nuclei at high energies \cite{z14}. It has been
shown that radiative corrections can simulate a variation of
 the magnetic
polarizability $\beta_{\pi}\simeq$ -0.2 from $\beta_{\pi}\simeq$ -5,
with plausible cuts for the outgoing photon.

\subsection{$\gamma\gamma \rightarrow \pi\pi$ at threshold}

Another way to get the pion polarizability is by means of
the measurement of $\gamma\gamma\rightarrow\pi\pi$ at threshold,
performed at  the new  Frascati $\Phi$-factory DA$\Phi$NE \cite{z15}.

A presentation of this new experimental facility has been
done already in the workshop and no further discussion on experimental
details has been done in this working group.

In summary the new  Frascati  $e^{+}e^{-}$ storage ring DA$\Phi$NE
 is supposed to deliver a luminosity $\simeq 5 \cdot 10^{32}$ cm$^{2}$
$s^{-1}$
at the $\Phi$ mass, with the possibility to increase the total
energy up to 1.5 GeV. Two detectors are under construction:
an all purposes detector, KLOE \cite{z16} mainly dedicated to CP
violation in $K$ decay, and FINUDA \cite{z17} mainly dedicated to hypernuclei
physics.

KLOE is expected \cite{z18} to detect $10^{3} \div 10^{4}$
times the events collected at present in
$e^{+}e^{-} \rightarrow e^{+}e^{-} \pi^{+}\pi^{-}$
and $e^{+}e^{-} \rightarrow e^{+}e^{-} \pi^{0}\pi^{0}$
at threshold.
Unfortunately at the $\Phi$ energy the decay $\Phi\rightarrow K_{S} K_{L}$
is an overwhelming
background (in $\sim$ 15 $\%$ of the events $K_{L}$ are not detected)
and tagging the outgoing $e^{+} e^{-}$ is needed.
 Two kind of tagging system for the outgoing leptons are foreseen.
  First of all there are two different rings for electrons and positrons
and the splitter magnet after the interaction region is a
suitable magnetic analyzer for the outgoing $e^{+} e^{-}$,
mostly forward emitted. Furthermore $e^{+} e^{-}$
emitted at larger angles are, in part,
detected by the central tracking detector in KLOE.
The $e^{+} e^{-}$ angular distribution depends on $m_{e}/E_{e}$ :
therefore there are more events at large angles in DA$\Phi$NE
respect to the high energy $e^{+} e^{-}$ storage rings.

By the way correlations in the azimuthal angles between the pions and the
outgoing leptons could be performed \cite{z19}, increasing the possibility
to disentangle the D wave contribution. Otherwise $\gamma\gamma$
interactions near threshold are supposed to provide mainly the $\pi\pi$ S wave,
which provides only $\alpha_{\pi} - \beta_{\pi}$.

 The overall double tagging efficiency is $\sim$ 15 $\%$ \cite{z18}.
The background from beam-beam bremmstrhalung is still under
study for evaluating the single tagging efficiency.

 In $\gamma\gamma$ interactions complications
related to any nuclear target are avoided, but it has been
demonstrated in the following theoretical discussion that the
extrapolation to the pion pole is much more difficult.
Therefore the conclusion has been achieved that
$\gamma\gamma$ interactions are not the best way
to get the pion polarizability. Nevertheless $\gamma\gamma$
interactions near threshold remain a very clean test of any theoretical
description of strong interactions at low energies.
\\

 Another possibility pointed out for measuring neutral and
charged $\alpha_{\pi}$ in
$e^{+}e^{-}$ is by means of $e^{+}e^{-} \to \pi\pi\gamma$
increased by the interference both with $\omega \to$ $\pi^{0}$ $\rho^{0}$ $\to$
$\pi_{0}$$\pi^{0}$$\gamma$ and, for the charged one,
also with radiative $\rho$ production.

\setcounter{subsection}{0}
\section{Theoretical issues in polarizability experiments\label{th}}

	 We begin with the process $\gamma\gamma\rightarrow\pi^0\pi^0$.
In this case the Born amplitude vanishes and the one-loop corrections
in Chiral Perturbation Theory (CHPT)\cite{wein79}-\cite{review} are finite
\cite{bico,dhlin}. The corresponding cross section
is independent of the free parameters of the chiral lagrangian and
does not agree with the experimental measurements at
Crystal Ball \cite{cball}, as well as with
calculations based on dispersion relations \cite{goble}-\cite{kaloshin},
even at low-energy.
The low-energy amplitude recently calculated to two-loops in CHPT
\cite{bgs} agrees with the Crystal Ball data and compares very well with the
results of a dispersive calculation by Donoghue and Holstein \cite{dohod}.

The value of the low-energy constants can be obtained in several ways, e.g.
by resonance exchange. The resonance saturation method provides empirical
values for the
scale-dependent renormalized constants of CHPT \cite{glan}-\cite{glnp}
         \be
L_i^r (\mu)=L_i^r ({\mu}_0)-\frac{{\Gamma}_i}{16\pi^2} ln\frac{\mu}{{\mu}_0}
{}~\co ~i=1,..,10
         \ee
with a scale $\mu$ in the range 0.5 $GeV$ -- 1 $GeV$ and a set of constants
$\Gamma_i$ defined in \cite{glan}-\cite{glnp}.
This method has been used in Ref. \cite{bgs} to pin down the couplings in the
$\gamma\gamma\rightarrow\pi^0\pi^0$ amplitude to order $p^6$.
In his talk J. Kambor discussed how to determine
these couplings from sum rules, exploiting the low- and high-energy behaviour.

Let us consider the vector-vector two-point function
\be
i\int d^4x e^{iqx}<0|T(V_{\mu}^a (x)V_{\nu}^b (0)|0> =
\delta^{ab} (q_{\mu}q_{\nu}-g_{\mu\nu}q^2 )\Pi^a (q^2) .
\ee
Following \cite{wein79}-\cite{glnp} and \cite{dohod} we write a
dispersion relation for $\Pi^a (q^2)$
\be
\Pi^a (q^2)=\frac{1}{\pi}\int ds\frac{Im\Pi^a (s)}{s-q^2-i\epsilon}
{}~+~subtractions \co
\label{na0}
\ee
in order to
make contact with the high-energy behaviour of the theory
desumed from the perturbative $QCD$ sum rules
\bea
\rho_V^a (s)& = &\frac{1}{\pi}Im\Pi^a (s)\co
         \nn
lim_{s\rightarrow\infty}\rho_V^a (s)& = &\frac{1}{8\pi^2}+O(\frac{1}{s}) ~,~
a=3,8
         \eea
showing that the spectral function $\rho_V^a (s)$ at high energy
goes like a constant plus higher-order terms that are suppressed
at least as $\frac{1}{s}$. Hence, the difference between two spectral
functions goes like $s$ at large $s$, and the integral
\be
\int ds\frac{\rho_V^3-\rho_V^8}{s}
\ee
converges. Also the once-subtracted dispersion relation for $\rho_V^3 (s)$
converges
\be
\int ds\frac{\rho_V^3}{s^2}~.
\ee

{}From the dispersion relation
\be
\Pi^3 (q^2)-\Pi^8 (q^2)=\frac{1}{\pi}\int ds
\frac{Im\Pi^3 (s)-Im\Pi^8 (s)}{s-q^2-i\epsilon}
\label{na1}
\ee
the sum rule is readily obtained
\be
\Pi^3 (0)-\Pi^8 (0)=\int ds
\frac{\rho_V^3-\rho_V^8}{s}~.
\label{na2}
\ee
Taking the $q^2$ derivative of the dispersion relations (\ref{na1})
and (\ref{na0}) yields the sum rules
\be
\frac{d}{dq^2}(\Pi^3 (0)-\Pi^8 (0))=\int ds
\frac{\rho_V^3-\rho_V^8}{s^2}
\label{na3}
\ee
and, respectively,
\be
\frac{d}{dq^2}\Pi^3 (0)=\int ds
\frac{\rho_V^3}{s^2}~.
\label{na4}
\ee
Notice that if one takes too many $q^2$-derivatives, then the integrals
become dominated by the threshold region.

As for the low-energy behaviour, J. Kambor showed how to use CHPT to calculate
$\Pi^a (q^2)$ for small $q^2$ values. This calculation is carried out in
$SU(3)\times SU(3)$ to the two-loop order, and the result depends on
the $O(p^4)$ and $O(p^6)$ low-energy constants $L_i$ and, respectively,
$d_j$ \cite{GK}. The integral on the r.h.s. of Eq. (\ref{na2}) can be
evaluated from the $e^+ e^-$ data.
In the narrow width approximation
one gets the following estimate \cite{KMS}:
\be
\int ds\frac{\rho_V^3-\rho_V^8}{s}=\frac{3}{4\pi\alpha^2}
\biggr( \frac{\Gamma_{\rho\rightarrow e^+ e^-}}{M_{\rho}}
-3\frac{\Gamma_{\omega\rightarrow e^+ e^-}}{M_{\omega}}
-3\frac{\Gamma_{\phi\rightarrow e^+ e^-}}{M_{\phi}}\biggr)
=(11.1\pm 2.0)\cdot 10^{-3} ~.
\ee
Thus, the integration region becomes
divided into three pieces, i.e. $4M_{\pi}^2\le s\le \Lambda_1$,
$\Lambda_1\le s\le\Lambda_2$, and $s\ge\Lambda_2$. Here we denote
by $\Lambda_{1,2}$ two cutoff values of about 0.4 $GeV$ and 2 $GeV$,
respectively, and $M_{\pi}$ is the pion mass. In the first region
the shape of the spectral function is obtained from the two-loop CHPT
calculation, i.e. $\rho_V =\rho_V^{1-loop}+\rho_V^{2-loop}$. In
the second (third) region the $e^+ e^-$ data (the perturbative $QCD$
calculation) can be used to obtain $\rho_V (s)$.
The result of this calculation (once it is completed) yields a scale
independent determination of $d_3$ contributing to
$\gamma\gamma\rightarrow\pi^0\pi^0$. J. Kambor expects an accuracy for
this estimate between 10 and 20 percent. This would be more precise than
the estimate carried out in \cite{KMS}
using a similar procedure (excluding, however, the two-loop contribution),
within the framework of the Generalized Chiral Perturbation Theory (GCHPT)
\be
d_3 = (9.4\pm 4.7)\cdot 10^{-6} ~.
\label{na5}
\ee

There are two more sum rules to consider. In particular Eq. (\ref{na3})
is effectively a sum rule for $L_9$ (the $d_i$ contributions to
$\Pi^3$ and $\Pi^8$ drop out of the sum rule), whereas Eq. (\ref{na4})
is a sum rule for $d_5$, $d_6$ contributing to
$\gamma\gamma\rightarrow\pi^+\pi^-$ and
$\gamma\pi^{\pm}\rightarrow\gamma\pi^{\pm}$. A similar treatment can be
applied to the sum rule for the axial-axial two-point function. This
gives an estimate of the low-energy constants $L_{1,2,3}$ contributing
to $\pi\pi$-scattering.

The GCHPT approach is described in \cite{reffks} (see also M. Knecht's
talk in the $\pi\pi$-scattering Working Group Section of these Proceedings).
Within this approach the cross section for
$\gamma\gamma\rightarrow\pi^0\pi^0$ and the pion polarizabilities
have been calculated up to and including $O(p^5 )$
\cite{KMS}. M. Knecht discussed the result of this calculation.
He showed that the cross section depends on the quark
mass ratio $r=\frac{2m_s}{m_u +m_d}$ and is consistent with the data
from Crystal Ball \cite{cball}, provided the value of the ratio is
at least a factor of two or three smaller than its standard value
in CHPT \cite{KMS}. M. Knecht showed that a
low-energy theorem analogous to the one
outlined above, but without taking into account the 2-loop contribution,
yields, through the evaluation of the dispersive integral via resonance
saturation, the following value for the $O(p^5 )$ constant $c$
defined in \cite{KMS}:
\be
c = -\frac{1}{r-1}(4.6\pm 2.3)\cdot 10^{-3} ~.
\label{na6}
\ee
In the standard CHPT case, i.e. for
\be
r = 2\frac{M_K^2}{M_{\pi}^2}-1=25.9 ~,
\label{na7}
\ee
the expression for $c$ given in Eq. (\ref{na6}) corresponds to
the value reported in Eq. (\ref{na5}). This is to be compared with
the value calculated from Appendix D of Ref. \cite{bgs}, using
resonance exchange
\be
d_3 = \pm 3.9\cdot 10^{-6} ~.
\ee

M. Pennington reviewed the dispersion relation treatment of the
$\gamma\gamma\rightarrow\pi^0\pi^0$ amplitude that represents
the data quite well \cite{pe}, \cite{pehan}. He showed how to
calculate the cross section from first principles, using a
relativistic and causal description based on the unitarity
of the scattering matrix. The prediction for low-energy
$\gamma\gamma\rightarrow\pi^0\pi^0$ and
$\gamma\gamma\rightarrow\pi^+\pi^-$ is based on the present
knowledge of the $\pi\pi$ phases and $PCAC$. M. Pennington
argued that precision measurement of
$\gamma\gamma\rightarrow\pi^0\pi^0,\pi^+\pi^-$ at the Frascati
$\Phi$-factory DA$\Phi$NE will restrict further the $\pi\pi$ phases
and determine where the chiral zero appears on-shell.

The electric and magnetic polarizabilities enter the
low-energy limit of the coupling
with the photon in the Compton amplitude
for any composite system. The dynamics of hadronic systems
can be probed by measuring the hadron
polarizabilities \cite{revpol}. In particular, the pion Compton scattering
can be investigated in this connection. The charged pion Compton amplitude
         \be
         \gamma(q_1)  \pi^+(p_1) \rightarrow \gamma(q_2)
         \pi^+(p_2)\co
         \ee
admits an expansion near threshold
          \bea
         T^C &=& 2 \left[ \vec{\epsilon}_1 \cdot \vec{\epsilon}_2 \,\! ^\star
 \left(
         \frac{\alpha}{\hmpp} - \bar{\alpha}_{\pi} \omega_1 \omega_2
         \right)- \bar{\beta}_{\pi} \left(\vec{q}_1 \times \vec{\epsilon}_1
         \right) \cdot \left(
         \vec{q}_2 \times \vec{\epsilon}_2 \,\! ^\star \right)
          + \cdots \right]
         \eea
         with $q_i^\mu = (\omega_i, \vec{q}_i)$.
Below we denote
\bea
(\alpha \pm \beta)^C &=&   \bar{\alpha}_{\pi} \pm \bar{\beta}_{\pi}
\co \nn
(\alpha \pm \beta)^N &=&   \bar{\alpha}_{\pi^0} \pm \bar{\beta}_{\pi^0} \; \; ,
\eea
for charged and neutral pions, respectively.

The charged pion polarizabilities have been determined in an experiment
on the radiative
pion-nucleus scattering $\pi^-A\rightarrow \pi^-\gamma A $ \cite{serpukov1}
and in the pion photoproduction process $\gamma p
\rightarrow \gamma \pi^+n$  \cite{lebedev}.
Assuming the constraint  $\apbc=0$
the two experiments yield\footnote{Throughout the following, we
express the values of the polarizabilities in units of $10^{-4} fm^3$ }
\bea
\ambc  =\left\{ \begin{array}{ll}
13.6 \pm 2.8 & \cite{serpukov1} \\
\; \; \; 40 \pm 24 & \cite{lebedev} \per
\end{array}
\right.
\label{cs3}
\eea
Relaxing the constraint
 $\apbc = 0$, one obtains from the Serpukhov data
\bea
 \apbc&=& \; \; 1.4 \pm 3.1 (\mbox{stat.}) \pm
2.5 (\mbox{sys.}) \; \;  \cite{serpukov2} \co \nn
 \ambc&=& 15.6 \pm 6.4 (\mbox{stat.}) \pm 4.4 (\mbox{sys.})
\; \; \cite{serpukov2} \per
\label{cs4}
\eea

At one-loop in CHPT one has \cite{lopol,z11,bbgm0}
\bea
\bar{\alpha}_{\pi^0}=-\bar{\beta}_{\pi^0}=
-\frac{\alpha}{96\pi^2M_{\pi}F^2} = -0.50 \; \; .
\label{cs6}
\eea
At order $O(p^6)$ it was calculated in Ref. \cite{bgs}
\bea
\bar{\alpha}_{\pi^0}&=&-0.35 \pm 0.10\co \nn
\bar{\beta}_{\pi^0}&=& \; \; \; 1.50 \pm 0.20.
\eea

The low-energy $\gamma \gamma \rightarrow \pi^+\pi^-$ data
\cite{dcharged} have been used in Ref. \cite{bbgm0} to obtain
information on $\bar{\alpha}_\pi$ and $\bar{\beta}_\pi$.
The result in \cite{bbgm0} yields the
numerical value for the leading-order
$\bar{\alpha}_\pi = 2.7 \pm 0.4$, plus systematic uncertainties due to
the $O(p^6 )$ corrections. The latter are not yet available.
A part of the corrections to the charged pion polarizabilities beyond the
one-loop order has been obtained in Refs. \cite{kaloshin86,ba2loop}
including the meson resonance contribution.

M. Knecht analyzed the $O(p^5)$ calculation from Ref. \cite{KMS}
of the $\pi^0$ polarizabilities
in GCHPT. The result $(\alpha +\beta)^N =0$ remains valid to this
order, whereas the remaining combination depends on the quark
mass ratio $r$. Hence this combination can have positive values
for $r$ much less than its standard CHPT value (\ref{na7}),
e.g. $(\alpha -\beta)^N =1.04\pm 0.60$ for $r=10$ \cite{KMS}.
For comparison we recall the standard CHPT prediction
for both combinations to the $O(p^6)$ order \cite{bgs}
\bea
(\alpha +\beta)^N &=& \; \; \; 1.15\pm 0.30 \co \nn
(\alpha -\beta)^N &=&-1.90\pm 0.20 .
\eea
The reason for the sign difference of $(\alpha -\beta)^N$ in GCHPT
with respect to the standard CHPT value has been traced back
by M. Knecht to a dominance by the positive $O(p^5)$ contribution
over the strongly suppressed negative
$O(p^4)$ contribution in GCHPT. This suppression is
related to a shift in the
position of the chiral zero, as $r$ becomes much smaller than the
standard CHPT value (\ref{na7}) \cite{KMS}.

Starting from the unitarized $S$-wave amplitudes
for neutral pions, M. Pennington displayed a proportionality
relation between $\ambn$ and the position of the chiral zero
$s_N$, showing that the former assumes values between -0.6
and -2.7 as the latter runs from $\frac{1}{2}M_{\pi}^2$ and
$2M_{\pi}^2$. He also discussed the validity of the errors
quoted in a recent estimate of
$\apbcn$ by Kaloshin and collaborators \cite{kalo}.
Here the polarizabilities appear as adjustable parameters
in the unitarized $D$-wave amplitudes, hence the values
of $\apbcn$ can be determined from the data
with the result \cite{kalo}
\bea
  \apbc &=& 0.22 \pm 0.06\; \; \cite{dcharged} \co \nn
 \apbn&=& 1.00 \pm 0.05 \; \;  \cite{cball} \per
\label{cs5}
\eea
M. Pennington, arguing on the partial wave analysis
of the data that shows large uncertainties even
at the $f_2$(1270) mass, concluded that the errors quoted in
(\ref{cs5}) for $\apbn$ are unbelievably small. His final conclusion,
that one must measure $\gamma\pi\rightarrow\gamma\pi$ can be
wholeheartedly shared, in view of future measurements of the
pion polarizabilities. In this respect, it is very important to devise
fully reliable methods that allow to extract the pion Compton scattering
amplitude from the measurement of the radiative pion photoproduction,
as discussed by the author.

\vspace{1.5cm}

\noindent
{\bf Acknowledgements}

It is a pleasure to thank the organizers of this Workshop for
a very stimulating working environment and the participants
to this working group for their precious help in preparing this report.
Very valuable help from R. Baldini in the preparation of this
work is also gratefully acknowledged.

\newpage
%\vspace{1.5cm}

\end{document}